%% file: main.tex
\documentclass[runningheads]{llncs}

\usepackage[T1]{fontenc}
\usepackage[utf8]{inputenc}
\usepackage{pgfplots}
\pgfplotsset{compat=1.14} 
\usepackage{url}
\usepackage{listings}
\usepackage{tikz}
\usetikzlibrary{shapes.geometric, arrows, shapes.misc, calc}

\tikzstyle{startstop} = [rounded rectangle, minimum width=3cm, minimum
height=1cm,text centered, draw=black, fill=red!10]
\tikzstyle{io} = [trapezium, trapezium left angle=70, trapezium right
angle=110, minimum width=3cm, minimum height=1cm, text centered, draw=black,
fill=green!10]
\tikzstyle{process} = [rectangle, minimum width=3cm, minimum height=1cm, text
centered, text width=3cm, draw=black, fill=yellow!10]
\tikzstyle{decision} = [diamond, aspect=2, minimum width=3cm, minimum
height=1cm, text centered, text width=5cm, draw=black, fill=blue!10, inner
ysep=-7pt]
\tikzstyle{arrow} = [thick,->,>=stealth]

\definecolor{graybg}{gray}{0.92}
\definecolor{graynumbers}{gray}{0.5}
\definecolor{bluekeywords}{rgb}{0.13,0.13,1}
\definecolor{greencomments}{rgb}{0,0.5,0}
\definecolor{redstrings}{rgb}{0.9,0,0}

\lstdefinestyle{CStyle}{
    backgroundcolor=\color{graybg},   
    commentstyle=\itshape\color{greencomments},
    keywordstyle=\bfseries\color{magenta},
    numberstyle=\tiny\color{graynumbers},
    stringstyle=\color{redstrings},
    basicstyle=\footnotesize\ttfamily,
    breakatwhitespace=true,         
    escapeinside={(*@}{@*)},
    breaklines=true,                 
    captionpos=b,                    
    keepspaces=true,                 
    numbers=left,                    
    numbersep=5pt,                  
    showspaces=false,                
    showstringspaces=false,
    showtabs=false,                  
    tabsize=2,
    frame=l,
    language=C
}

\lstdefinestyle{PyStyle}{
    backgroundcolor=\color{graybg},   
    commentstyle=\itshape\color{greencomments},
    keywordstyle=\bfseries\color{magenta},
    numberstyle=\tiny\color{graynumbers},
    stringstyle=\color{redstrings},
    basicstyle=\footnotesize\ttfamily,
    breakatwhitespace=true,         
    escapeinside={(*@}{@*)},
    breaklines=true,                 
    captionpos=b,                    
    keepspaces=true,                 
    numbers=left,                    
    numbersep=5pt,                  
    showspaces=false,                
    showstringspaces=false,
    showtabs=false,                  
    tabsize=2,
    frame=l,
    language=python
}

\lstdefinestyle{TermStyle}{
    backgroundcolor=\color{lightblue},   
    commentstyle=\itshape\color{greencomments},
    keywordstyle=\bfseries\color{white},
    numberstyle=\tiny\color{graynumbers},
    stringstyle=\color{redstrings},
    basicstyle=\footnotesize\ttfamily,
    breakatwhitespace=true,         
    escapeinside={(*@}{@*)},
    breaklines=true,                 
    captionpos=b,                    
    keepspaces=true,                 
    numbers=left,                    
    numbersep=5pt,                  
    showspaces=false,                
    showstringspaces=false,
    showtabs=false,                  
    tabsize=2,
    frame=l,
    language=C
}

\begin{document}

\title{An OpenMP translator for the GAP8 MPSoC}
\subtitle{\textit{Position Paper}}
\author{Reinaldo Agostinho de Souza Filho\inst{1} \and
Diego V. Cirilo do Nascimento\inst{2} \and
Samuel Xavier-de-Souza\inst{1}}
\authorrunning{R. A. de Souza Filho et al.}

\institute{Universidade Federal do Rio Grande do Norte, Natal-RN, Brazil \\ 
\email{\{reinaldo.souza,samuel\}@dca.ufrn.br} \and
Instituto Federal do Rio Grande do Norte, Natal-RN, Brazil\\
\email{diego.cirilo@ifrn.edu.br}
}

\maketitle

\begin{abstract}
  One of the barriers to the adoption of parallel computing is the inherent
  complexity of its programming. The Open Multi-Processing (OpenMP) Application
  Programming Interface (API) facilitates such implementations, providing high
  abstraction level directives. On another front, new architectures aimed at
  low energy consumption have been developed, such as the Greenwaves
  Technologies GAP8, a Multi-Processor System-on-Chip (MPSoC) based on the
  Parallel Ultra Low Power (PULP) Platform. The GAP8 has an 8-core cluster and
  a Fabric Controller(FC) master core. Parallel programming with GAP8 is very
  promising on the efficiency side, but its recent development and lack of a
  robust OS to handle threads and core scheduling complicate a simple
  implementation of the OpenMP APIs. This project implements a source to source
  translator that interprets a limited set of OpenMP directives, and is capable
  of generating parallel microcontroller code manipulating the cores directly.
  The preliminary results obtained in this work shows a reduction of the code
  size, if compared with the base implementation, proving the efficiency of the
  project to ease the programming of the GAP8. Further work is need in order to
  implement more OpenMP directives.

\end{abstract}

\section[Introduction]{Introduction}
\label{sec:introduction}

Over 17 billion Internet of Things (IoT) devices were connected in 2018
\cite{iotgrow}. Such devices have tight energy budgets, due to dimensions,
battery size and heat dissipation constraints.
Efficiency is the key to overcome these limitations while providing the
required performance, demanding efforts from both the hardware and the software
fronts. 

New Instruction Set Architectures (ISAs), such as the RISC-V were created with
energy efficiency as a goal \cite{risc-v}, as well as new devices as the
Greenwaves Technologies GAP8, a Multi-Processor System on Chip (MPSoC) based on
the Parallel Ultra Low Power (PULP) platform \cite{gap_aval,gap8}.

The PULP Platform started as a joint effort between the Integrated Systems
Laboratory (IIS) of Eidgenössische Technische Hochschule (ETH) Zürich and
Energy-efficient Embedded Systems (EEES) group of the University of Bologna in
2013 to explore new and efficient architectures for ultra-low-power processing. 
The aim is to develop an open source, scalable hardware and software research
and development platform with the goal to break the energy efficiency barrier
within a power cost of a few milliwatts, as well as satisfy the computational
demands of IoT applications requiring flexible processing of data \cite{pulp}.

Although it stands out as an innovative technology, one of the limiting factors
to the popularization of GAP8 is the complexity of its parallel programming.
At the time of this writing, there is no support to more complex operating
systems, capable of managing parallel operations.
Presently GAP8 supports three Real Time Operating System (RTOS): Mbedos,
PULPOS and FreeRTOS. None is able to assign threads to specific hardware cores
using functions such as pthreads. The Software Development Kit (SDK) of
GAP8 supports the C and C++ programming languages.

OpenMP is an open source API developed by the OpenMP Foundation that aims to
offer paralelization directives in a higher abstraction level than the usual
native APIs, reducing the complexity of implementation of parallel applications.

The goal of this project is to develop an OpenMP source to source
translator for the GAP8, simplifying the development of parallel applications
for the architecture. 

Currently, this translator implements the following OpenMP directives, clauses
and functions: 
\begin{multicols}{2}
\begin{enumerate}
    \item \texttt{parallel}
    \item \texttt{for}
    \item \texttt{parallel for}
    \item \texttt{critical}
    \item \texttt{single}
    \item \texttt{num\_threads}
    \item \texttt{reduction}
    \item \texttt{private}
    \item \texttt{shared}
    \item \texttt{omp\_get\_thread\_num()}
    \item \texttt{omp\_get\_num\_threads()}
\end{enumerate}
\end{multicols}

\section{Parallel Programming in the GAP8}
\label{sec:cap2}
\begin{figure}
    \centering
    \includegraphics[width=0.9\textwidth]{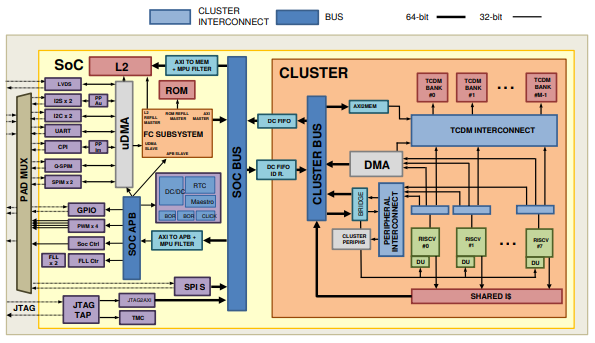}
    \caption{GAP8 architecture \cite{gap_manual}.}
    \label{fig:gap8-arch}
\end{figure}

GAP8 is an open-source IoT processor that implements an extended
version of the RISC-V instruction set. It is composed of one Fabric Controller
(FC) core, and eight cores in a cluster. Each core on the cluster has its own
data cache.
GAP8 is uniquely optimized to perform high-performance imaging and audio
processing algorithms including convolutional neural network inference with
extreme energy efficiency \cite{gap_manual}. The Figure \ref{fig:gap8-arch}
shows the architecture of the GAP8.

A Software Development Kit (SDK) is provided by the manufacturer, and allow the
programming of the system in C or C++. Besides the usual peripheral and setup
functions, functions for controlling the parallel operations are provided \cite{gap_SDK}.

\subsection{Cluster usage}
The RTOS runs on the FC core, and parallel operations must be offloaded to the
cluster. The cluster is able to run on an independent voltage and frequency
domain, and is turned off by default, aiming to reduce the power consumption.

The first step is to issue the \texttt{CLUSTER\_Start(0, number\_of\_cores)}
function power-up the cluster.
\texttt{CLUSTER\_SendTask(0, Master\_entry, (void*) data, 0)} will send the
task to the cluster.

This task is a previously defined on the master function with arguments (data)
that must be a void pointer and return to the original value later with a cast
for the desired type all that is sent to the cluster with a function to fork
to the cores on cluster.

Within the \texttt{Master\_entry} the function
\texttt{CLUSTER\_CoresFork(function, data)} is called, and that basically
replicates the function that must also be previously created for the cluster
passing data (or as you would like to call its argument) as argument.

The data that is passed by reference, using structures or vectors to access all
the necessary arguments.

Another problem would be not being able to reset the \texttt{Master\_Entry}
which limits multiple parallel regions to do different functionality, having to
create n master functions for n regions besides the n functions to be fired.

\subsection{Memory allocation}

Private data can be directly passed to the cluster using
\texttt{L1\_Malloc(size)} so the cluster elements can access directly from
their own memory then need to issue a \texttt{L1\_Free(size)} to  clean the L1
memory address data from the cluster allowing to use it again without false
fully memory of the program.  This post-allocation data should be sent as
previously said by the master to the functions that should handle them.

If you need to have a variable that needs to be accessed by the FC and the
cluster, it must be global.

\subsection{Shutting down the Cluster}

To terminate the cluster work you must use the \texttt{CLUSTER\_Stop()}
function, and have the option to use \texttt{CLUSTER\_Wait()} before to ensure
that all cores finish before stopping the group.
It is worth remembering that \texttt{L1\_Free} should be used before stopping
the cluster otherwise there will not be another chance.

\section[The basics of the translator]{The basics of the translator}

\begin{figure}
    \centering
    \scalebox{0.6}{\input{fig/translator-workflow}}
    \caption{Translator workflow.}
    \label{fig:parser}
\end{figure}
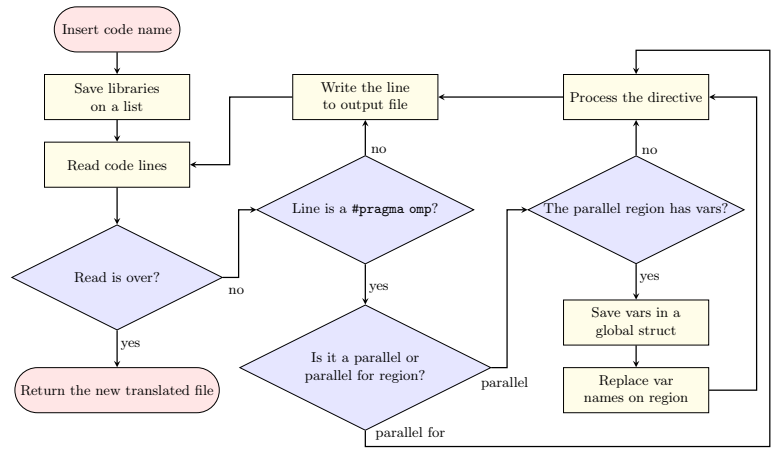

A C code is passed as argument to the
code \texttt{rei\_omp.py}. The code is sent to an output folder that must be
created before execution with the name \texttt{output}. The code created has the same name as the one passed
as parameter added with \texttt{\_gap} before the \texttt{.c}. A Diagram
is show on Figure \ref{fig:parser}.

\section[Tests and Results]{Tests and Results}
\label{sec:cap5}
For verifying the operation of the developed code, tests were selected from the
source codes of the book of Peter Pacheco Parallel computing \cite{pacheco}.

All code requests input and output, for a embedded device this is not reality
although there is a version of \texttt{printf} made for GAP\_SDK it is really
necessary to adapt the code to work on a embedded device, respecting the idea
of algorithms and all their directives.
The ratio of the original program to the generated program is considerable, the
programs had very similar or equal results in the tests, we use the following
programs.

\begin{enumerate}
    \item \texttt{omp\_hello.c} -- A ``Hello World'' in parallel;
    \item \texttt{omp\_pi.c} -- The calculation of $\pi$;
    \item \texttt{omp\_trap2a.c} -- Trapezoidal rule integration.
\end{enumerate}

In addition to the \texttt{hello\_for.c}and \texttt{hello\_cru.c} codes of our
authorship based on own gap\_sdk programs

The number of rows of the original OpenMP codes compared to the generated codes
can be seen in the following table:

\begin{table}
    \centering
    \caption{Code lines comparison}
    \begin{tabular}{||c|c|c|c||}
    
    \hline
          Source & Original number of lines & Converted code lines & Reduction
          percentage  \\
    \hline
          \texttt{omp\_pi.c} & 40 & 135 & 70\% \\
          
    \hline
          \texttt{omp\_trap2a.c} & 98 & 265 & 63\%\\
    \hline
          \texttt{omp\_hello.c} & 22 & 79 & 72\%\\
    \hline
          \texttt{hello\_for.c} & 24 & 121 & 80\% \\
    \hline
          \texttt{hello\_cru.c} & 49 & 266 & 81\%\\
    \hline 
    \end{tabular}
    \label{tab:line_code}
\end{table}

omp\_pi.c estimates de value of pi, omp\_trap2a.c calculates an integral with
the trapezoidal rule, omp\_hello is a simple parallel hello world, hello\_for
is a program used to debug exclusively the for loop, hello\_cru.c combines all
directives and clauses the last two were made by purpose of tests and the
others Parallel Programming adapted codes to run in GAP8(excluding inputs and
adding libraries) to validate the project.

It is noticeable the decrease of writing work of the programmer, since the
functionality keeps considering the restrictions of a System on chip (SoC) we
have a gain of time that can be valuable for a programmer.

\section[Conclusion]{Conclusion}
\label{sec:conclusion}

With the regular expressions use it was possible to implement a simplified
version of OpenMP to run on GAP8, even with the limitations of systems that
deal with MPSoCs.  GAP8 has a potential for helps the industry to grow and
evolve.  Its energy savings can be a solution to the exponential growth of
IoTs, more and more devices can benefit from parallel computing without
worrying about limiting energy consumption and heat dissipation. Besides that
GAP8 provides a offloading from the cluster improving the expected results.

The need to grow technologically is inevitable for the population, which is
always seeking information solutions and more and more data are produced.
OpenMP can make the development of embedded projects advance and possibly
popularize the GAP8 architecture.  Translation code level is reached on this
work, the goal is to reach the compilation level in addition to include other
OpenMP directives. It is hoped that community collaboration will allow this
project to grow\footnotemark[1].  Possible inclusion projects of OpenMP
interpreters can be included in the compilers of all future platforms.

\footnotetext[1]{The source code and tests are available on the project's Github
repository at:\\ \url{https://github.com/cevero/omp-gapuino}}

The results of the translator usage shows a reduction of the codes size, claim
less time to a programmer to worry about GAP8 singleness, and enjoying a most
known API usage to develop their projects.

This is a work in progress, even with the limitations presented it was possible
to notice a considerable decrease in the workload of the programmer. Continuing
with this project could give the entire community a contribution and for all
those who wish to innovate with performance and low consumption.  The target of
this job will focus on work in more directives and clauses and reach a
compilation level.  Potentially, it may increase the adhesion of PULP platform
and derivatives projects.

\bibliographystyle{splncs04}
\bibliography{bibliography}

\end{document}

%% file: fig/translator-workflow.tex
\begin{tikzpicture}[node distance=2cm]

\node (start) [startstop] {Insert code name};
\node (p1) [process, below of=start, yshift=0.5cm] {Save libraries on a list};
\node (p2) [process, below of=p1, yshift=0.5cm] {Read code lines};
\node (d1) [decision, below of=p2, yshift=-0.5cm] {Read is over?};
\node (p3) [process, right of=p1, xshift=3.5cm] {Write the line to output file};
\node (d2) [decision, below of=p3, yshift=-0.5cm] {Line is a \texttt{\#pragma
  omp}?} ;
\node (d4) [decision, right of=d2, xshift=4cm] {The parallel region has vars?};
\node (p8) [process, above of=d4, yshift=0.5cm] {Process the directive};
\node (d3) [decision, below of=d2, yshift=-1.5cm] {Is it a parallel or parallel
  for region?};
\node (p6) [process, below of=d4, yshift=-0.5cm] {Save vars in a global struct};
\node (p7) [process, below of=p6, yshift=0.5cm] {Replace var names on region};
\node (stop) [startstop, below of=d1, yshift=-0.5cm] {Return the new translated
  file};

\draw [arrow] (start) -- (p1);
\draw [arrow] (p1) -- (p2);
\draw [arrow] (p2) -- (d1);
\draw [arrow] (d1.east) node[anchor=north,xshift=0.3cm,yshift=-0.1cm] {no} --
  ++(10pt,0) |- (d2.west);
\draw [arrow] (d1.south) node[anchor=north,xshift=0.3cm,yshift=-0.1cm] {yes} --
  (stop);
\draw [arrow] (d2.north) node[anchor=north,xshift=0.3cm,yshift=0.3cm] {no} --
  (p3.south);
\draw [arrow] (p3.west) -| ($(p2.east)+(25pt,0)$) -- (p2.east);
\draw [arrow] (d2.south) node[anchor=north,xshift=0.3cm,yshift=-0.3cm] {yes} --
  (d3.north);
\draw [arrow] (d3.east) node[anchor=north,xshift=0.3cm,yshift=-0.1cm]
  {parallel}-- ++(10pt,0) |- (d4.west);
\draw [arrow] (d3.south) node[anchor=north,xshift=1cm,yshift=0.2cm]
  {parallel for}-- ++(0,-10pt)  -- ++(255pt,0)-- ++(0,250pt) -| (p8.north);
\draw [arrow] (p8.west) -- (p3.east);
\draw [arrow] (d4.south) node[anchor=north,xshift=0.3cm,yshift=-0.1cm] {yes} --
  (p6);
\draw [arrow] (d4.north) node[anchor=north,xshift=0.3cm,yshift=0.3cm]{no} --
  (p8.south);
\draw [arrow] (p6) -- (p7);
\draw [arrow] (p7.east)  -- ++(30pt,0) |- (p8.east);

\end{tikzpicture}